\documentclass[showpacs,prl,superscriptaddress,nofootinbib, twocolumn]{revtex4}
\usepackage{graphicx}
\usepackage{color}
\usepackage{amssymb,amsfonts,amsmath}
\usepackage{natbib}
\usepackage{epstopdf}
\usepackage{subfigure}

\begin{document}

\title{Noise driven neuromorphic tuned amplifier}

\author{Duccio Fanelli}  \affiliation{Dipartimento di Fisica e Astronomia and CSDC, Universit\`{a} degli Studi di Firenze, via G. Sansone 1, 50019 Sesto Fiorentino, Italia}
\affiliation{INFN Sezione di Firenze, via G. Sansone 1, 50019 Sesto Fiorentino, Italia}
\author{Francesco Ginelli} 
\affiliation{SUPA, Institute for Complex Systems and Mathematical Biology, Kings College, University of Aberdeen, Aberdeen AB24 3UE, United Kingdom}
\author{Roberto Livi} 
\affiliation{Dipartimento di Fisica e Astronomia and CSDC, Universit\`{a} degli Studi di Firenze, via G. Sansone 1, 50019 Sesto Fiorentino, Italia}
\affiliation{INFN Sezione di Firenze, via G. Sansone 1, 50019 Sesto Fiorentino, Italia}
\author{Niccol\'o Zagli} 
\affiliation{Dipartimento di Fisica e Astronomia and CSDC, Universit\`{a} degli Studi di Firenze, via G. Sansone 1, 50019 Sesto Fiorentino, Italia}
\author{Clement Zankoc} 
\affiliation{Dipartimento di Fisica e Astronomia and CSDC, Universit\`{a} degli Studi di Firenze, via G. Sansone 1, 50019 Sesto Fiorentino, Italia}
\affiliation{INFN Sezione di Firenze, via G. Sansone 1, 50019 Sesto Fiorentino, Italia}

\maketitle

{\bf  Living systems implement and execute an extraordinary plethora of computational tasks \cite{alberts, sole2}. The inherent degree of large scale coordination emerges as a global property, from the intricate sea of microscopic interactions. The brain, with its structural and functional architecture, represents an emblematic example of hierarchic self-organization \cite{kandel}:  elementary units, the neurons,  act much like instruments of an orchestra, which combine diverse timbres to create harmonious symphonies. Neurons come indeed in different types, varying in shapes, connections and electrical properties. They all team up to process external stimuli from a number of sources and integrate the information to yield, from neurons to mind, different cognitive faculties. Identifying the coarse grained modules that exert, from bottom to up, pivotal neuronal functions constitutes a goal of paramount importance.
On the other hand, the brain and its unraveled secrets could inspire novel biomimetics technologies to adaptively handle complex problems. Here, we investigate the intertwined stochastic dynamics of two populations of excitatory and inhibitory units, arranged in a directed lattice \cite{bressloff, ButlerPNAS}. The endogenous noise \cite{vankampen,mckanenewman, dauxois} seeds a coherent amplification across the chain generating giant oscillations with tunable frequencies, a process that the brain could exploit to enhance, and eventually encode, different signals. The system works as an out-equilibrium thermal device under stationary operating conditions, the associated entropy production rate being analytically quantified \cite{Schnakenberg,thermo, thermo1}. The same scheme could be invoked to design a novel family of manmade detectors capable of reacting to spatially distributed low intensity alerts.}

Simple deterministic models can be designed so to exemplify, at the mesoscopic level, the prototypical evolution of excitatory and inhibitory units, 
organized in two mutually competing populations. This is for instance the case of the celebrated Wilson-Cowan (WC) model of excitatory and inhibitory 
neurons \cite{WC1,WC2}. As opposed to mechanistic approaches targeted to single neuronal functioning, the WC model and its numerous variants indulge on a coarse grained description of the examined neuronal dynamics. 
Averaged families of homologous constituents are introduced and made to interact via non linear mean-field couplings, which epitomize the mechanism of 
threshold activation as displayed by individual neurons. By replicating the WC model on each patch of a supposedly heterogeneous network, and assuming that local activation gets modulated 
by adjacent populations, yield a spatially extended framework where coherent patterns of activation can organize and flow. Stochastic perturbation can however play a remarkable contribution \cite{Goychuck, Negahbani, Cowan2016, Wallace2011}. In particular, finite size effects manifest as an endogenous source of disturbance, which ultimately reflects the inherent discreteness of the scrutinized medium \cite{bressloff,vankampen,mckanenewman,bartlett}.  Building on these premises,  we will consider a minimal model for discrete agents in mutual interaction via excitatory and inhibitory loops \cite{Zankoc}. The model is formulated in terms of a birth/death stochastic process. In the idealized deterministic limit, this latter converges to a set of rate equations for the densities of active excitatory and inhibitory neurons, reminiscent of the WC type.  Endogenous noise instigates robust correlations across the lattice, and fuels the process of tuned amplification that we shall hereafter illustrate (see Figure \ref{fig2} and the annexed movies). 

\begin{figure}
 \centering
   {\includegraphics[width=7.5cm]{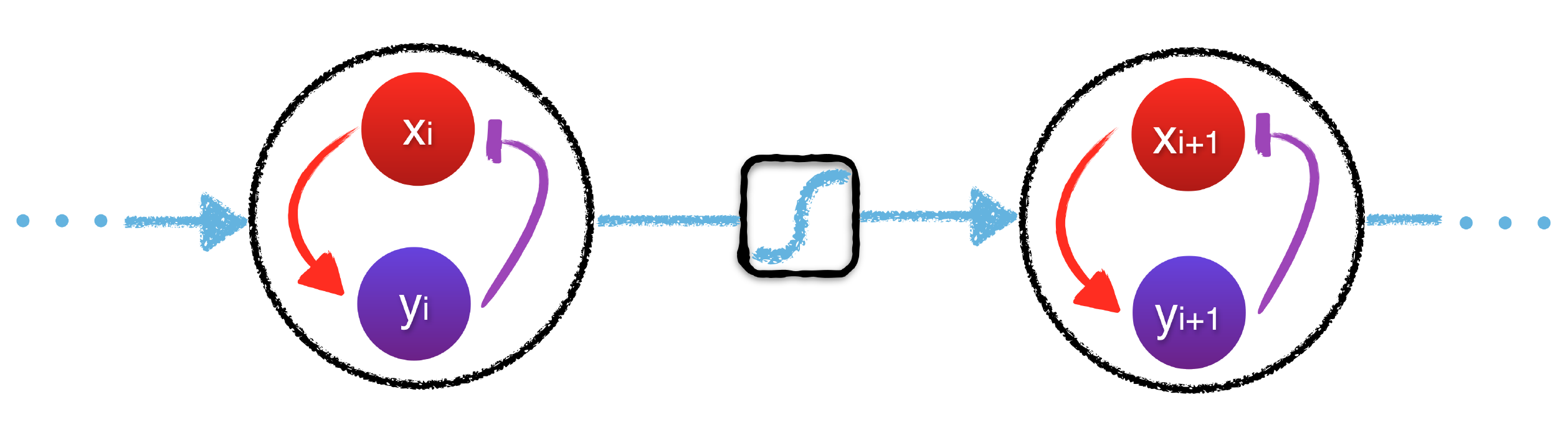}}
   \caption{Schematic layout of the neuromorphic circuit.}
   \label{fig1}
  \end{figure}

\begin{figure}
 \centering
 %\subfigure[]
   {\includegraphics[width=6.5cm]{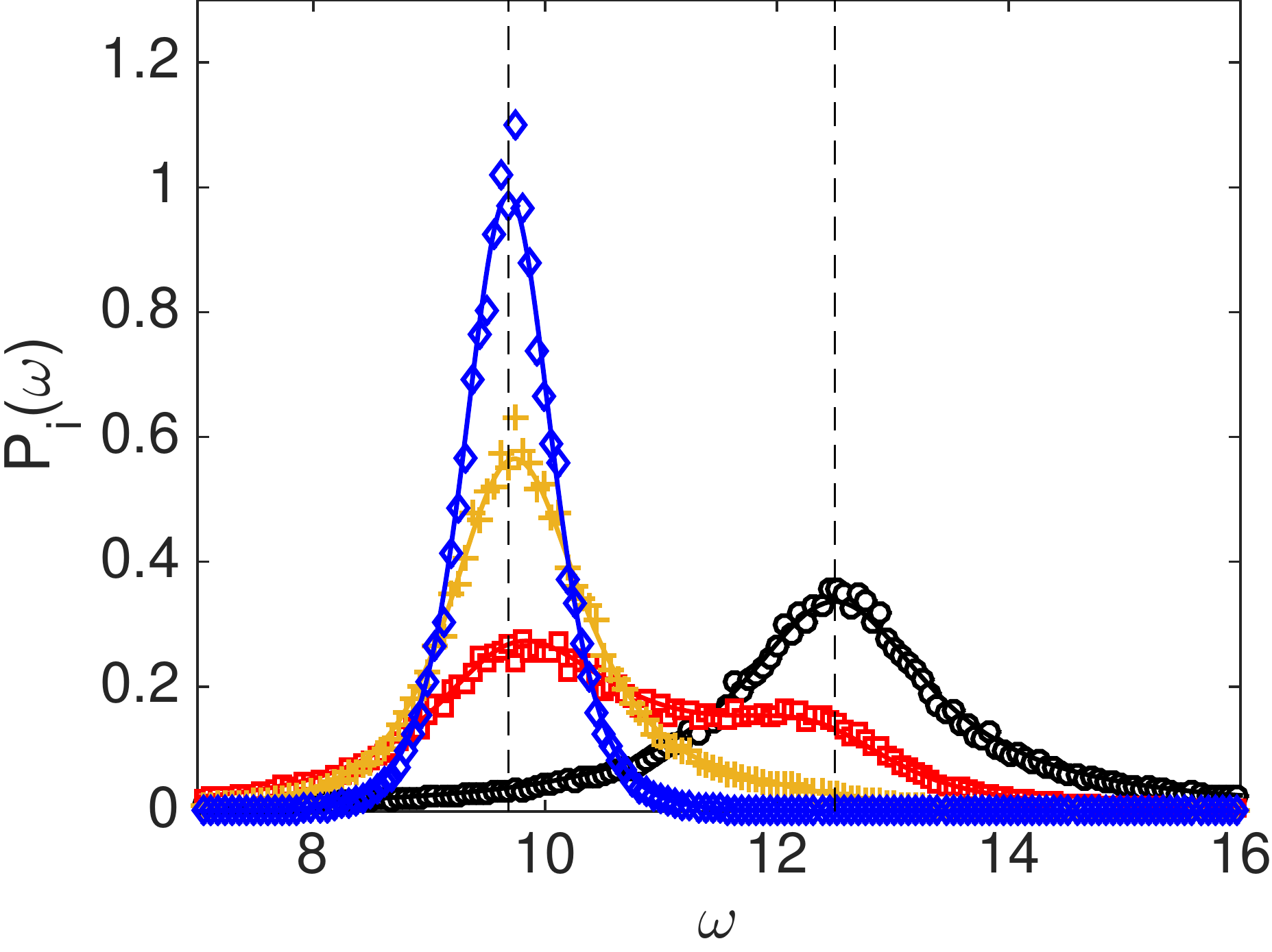}}
 %\hspace{5mm}
  %\centering
 %\subfigure[]
   {\includegraphics[width=6.5cm]{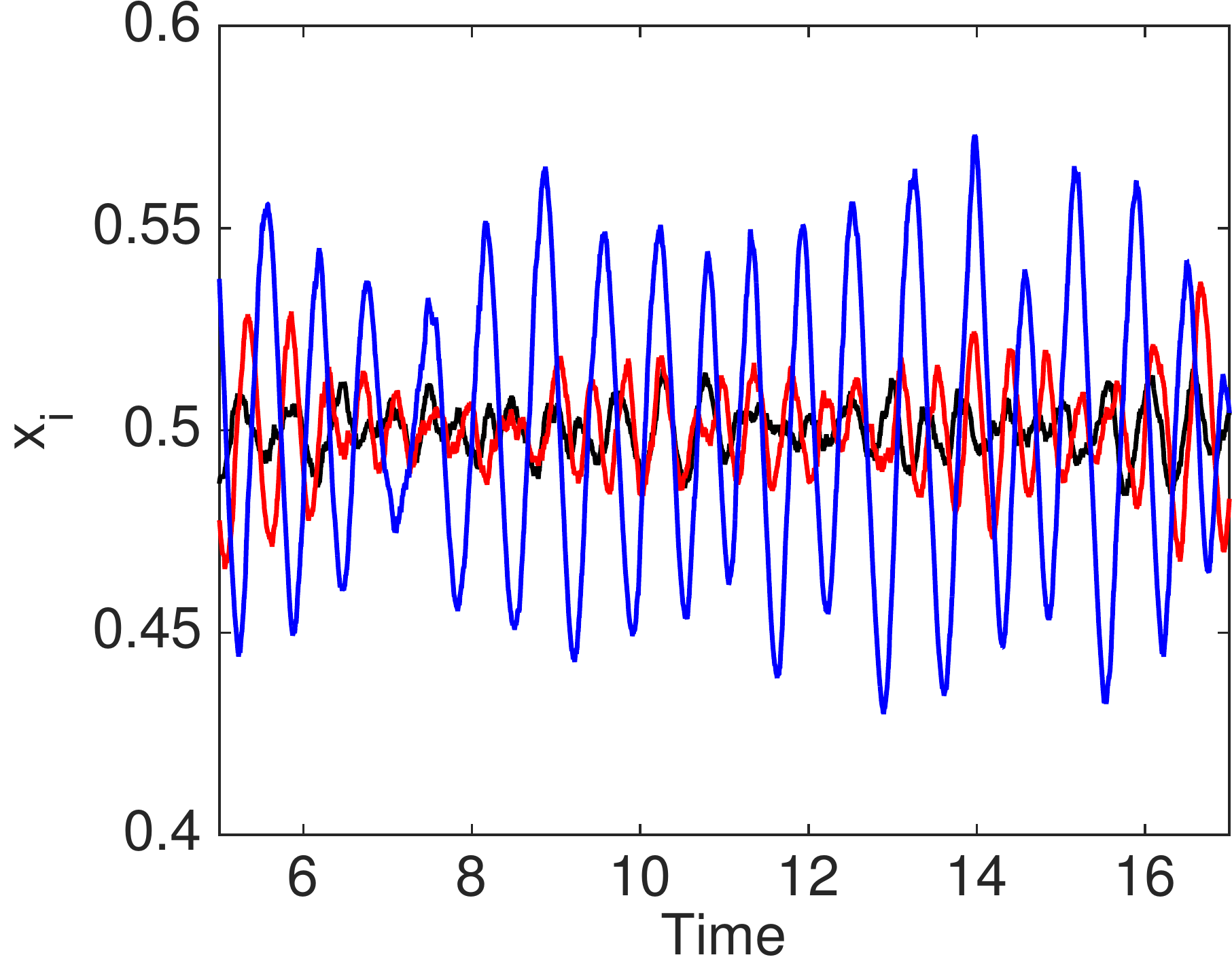}}
 %\hspace{5mm} 
  %\hspace{5mm}
   \caption{Upper panel: the theoretical power spectrum $P_{i}(\omega)$ (\ref{PS}) for  excitators ($X$), for $i=1,..,4$ is plotted with a solid line. Symbols refer to the power spectra computed from averaging independent realizations of the stochastic dynamics (\ref{LE}). In the SI we show that the nonlinear Langevin dynamics (\ref{LE}), returns results which are in excellent agreement with exact stochastic simulations based on on Gillespie algorithm \cite{gillespie}. The rightmost vertical dashed line is traced at $\omega_0$, the leftmost at $\omega_1$.  Here $r=50$, $D=10$ and $V=10^6$. Lower panel: stochastic trajectories on different nodes. Noisy self-sustained oscillation of modest amplitude are displayed on the first node of the lattice (black line). The amplitude of the oscillations grows steadily across the chain (red line on node $2$ and blue line on node $8$) and become progressively more regular.}
   \label{fig2}
  \end{figure}

Label $X_i$  (resp. $Y_i$), with $i\!=\!1,2,\ldots,\Omega$, the excitatory (resp. inhibitory) agents belonging to the mean-field interacting
patch (node) $i$ of volume $V_i$. The patches are organized in a one dimensional lattice, as depicted in Figure \ref{fig1}, with directional couplings. 
Individual elements are subject to the following birth and death chemical reactions: 

\begin{equation*}
\begin{array}{lclcl}
\emptyset & \overset{f[s_{x_i}]}{\longrightarrow} & X_i \qquad  \emptyset & \overset{f[s_{y_i}]}{\longrightarrow} & Y_i \\
X_i & \overset{1}{\longrightarrow} \emptyset  & \qquad Y_i & \overset{1}{\longrightarrow} & \emptyset \nonumber
\end{array}
\label{chemical}
\end{equation*}

where $\emptyset$ denotes an infinite reservoir; $f(s)\!=\!1/(1\!+\!\exp (\!-\!s))$ is a sigmoid function which captures the saturating response of neurons to
external stimuli \cite{WC1,WC2, bressloff}. Further:   

\begin{eqnarray*}
%s_{x_i}  &&\!=\!   \!-\! r \left( \frac{n_{Y_i}}{V} \!-\! \frac{1}{2} \right)+{D \left( \frac{n_{X_{i}}}{V} \!-\! \frac{n_{X_{i-1}}}{V} \right)-D \left( \frac{n_{Y_{i}}}{V} \!-\! \frac{n_{Y_{i-1}}}{V} \right)} \\ 
%s_{y_i}  &&\!=\!    r \left( \frac{n_{X_i}}{V} \!-\! \frac{1}{2} \right)+{D \left( \frac{n_{X_{i}}}{V} \!-\! \frac{n_{X_{i-1}}}{V} \right)-D \left( \frac{n_{Y_{i}}}{V} \!-\! \frac{n_{Y_{i-1}}}{V} \right)} 
s_{x_i}  &&\!=\!   \!-\! r \left(y_i\!-\! \frac{1}{2} \right)+{D \left( x_{i-1} \!-\! x_{i}\right)-D \left(y_{i-1}\!-\! y_{i} \right)} \\ 
s_{y_i}  &&\!=\!     r \left(x_i\!-\! \frac{1}{2} \right)+{D \left( x_{i-1} \!-\! x_{i}\right)-D \left(y_{i-1}\!-\! y_{i} \right)}
\end{eqnarray*}

where $x_i=n_{X_i}/V_i$ (resp. $y_i=n_{Y_i}/V_i$) is the concentration of excitators (resp. inhibitors) on node $i$; $n_{X_i}$ and $n_{Y_i}$
respectively identify the number of elements of type $X$ and $Y$ on patch $i$.  The  intensity of the couplings between adjacent populations is set by the positive defined parameter $D$. The 
directed nearest neighbors interaction is mediated by the non linear filter $f$, a scheme assumed in the literature to plausibly represent synaptic interactions. $r>0$ acts as a control parameter of the local dynamics. 
In the uncoupled limit ($D=0$), a large population of inhibitors damps the corresponding population of excitators, on the same patch. Similarly,  a large excitatory population stimulate a local growth of the inhibitors.   

Introduce $P({\boldsymbol v},t)$ to label the probability for the system to be in state ${\boldsymbol v } \!=\! ({x_1},{y_1},{x_2},{y_2}, \ldots, {x_\Omega},{y_\Omega})$ at time $t$. Transitions from one state to another are dictated by the above chemical equations. $T({\boldsymbol v}' | {\boldsymbol v})$ stands for the transition rate from state  ${\boldsymbol v} $ to state ${\boldsymbol v} '$, compatible with the former. The dynamics of the 
system is governed by a master equation \cite{vankampen, gardiner} which can be cast in the generic form
%\begin{equation}
$\frac{\partial}{\partial t} P({\boldsymbol v},t) \!=\! \sum_{\boldsymbol v'} T \left( {\boldsymbol v} | {\boldsymbol v'} \right) P({\boldsymbol v'},t)  \!-\! T \left( {\boldsymbol v} '| {\boldsymbol v }\right) P({\boldsymbol v},t)$.
%\label{ME}
%\end{equation}

One can then seek to approximate the exact master equation via the Kramers Moyal expansion \cite{gardiner, rogers, biancalani1, biancalani2, biancalani3} assuming large enough $V_1$, and $\gamma_i=V_i/V_1 = \mathcal{O}(1)$. Performing the calculation as detailed in the Supplementary Information (SI) yields:

\begin{eqnarray}
\label{LE}
\frac{d}{d\tau}{x}_i&=&\frac{1}{\gamma_i}\big[f(s_{x_i})-{x}_i  \big]+\frac{1}{\gamma_i \sqrt{V_1}}\sqrt{{x}_i+f(s_{x_i})}\lambda_i^{(1)} \\
\frac{d}{d\tau}{y}_i&=&\frac{1}{\gamma_i}\big[f(s_{y_i})-{y}_i  \big]+\frac{1}{\gamma_i \sqrt{V_1}}\sqrt{{y}_i+f(s_{y_i})}{\lambda}_i ^{(2)} \nonumber
\end{eqnarray}

where ${\boldsymbol \lambda} \!=\! ({\lambda_1^{(1)}},{\lambda_1^{(2)}},{\lambda_2^{(1)}},{\lambda_2^{(2)}}, \ldots, {\lambda_{\Omega}^{(1)}},{\lambda_{\Omega}^{(2)}})$ is a Gaussian stochastic variable with zero mean and correlator $\langle \lambda_i^{(l)} \lambda_j^{(m)} \rangle = \delta_{ij} \delta_{lm} \delta(\tau-\tau')$. Here, $\tau=t/V_1$. In the thermodynamic limit $V_1 \rightarrow \infty$, the above stochastic
equations reduce to a deterministic system which admits the homogenous fixed point  $x^{*} =y^{*} = 1/2$, $\forall i$. The stability of the fixed point can be assessed by computing the eigenvalues of  the $2 \Omega \times 2 \Omega$ Jacobian matrix ${\boldsymbol J}$, that we shall explicitly report in the SI. We will begin by assuming nodes of identical capacity $V_i=V$, which entails $\gamma_i=1$.  The spectrum of ${\boldsymbol J}$ is hence degenerate, owing to the peculiar structure of the  block diagonal Jacobian matrix. Two eigenvalues read  $\lambda_{1,2}=-1 \pm i r/4$ and coincide with those obtained when working with just one patch ($\Omega=1$).  The other eigenvalues are $\lambda_{3,4} = \left[-1 \pm \sqrt{\left(\frac{r}{8}(D-\frac{r}{2})\right)} \right]$, each with multiplicity $\Omega-1$. As expected, $\lambda_{3,4}$ converge to $\lambda_{1,2}$, when $D \rightarrow 0$.  Based on this relation, we can immediately conclude that the homogeneous fixed point is linearly stable provided $D<D_c \equiv \frac{r}{2} + \frac{8}{r}$. Importantly,  the eigenvalues are complex for $D<r/2<D_c$, an observation that plays a crucial role for what it follows. Summing up, for $D<r/2$, the deterministic system displays, at equilibrium, a uniform level of activity,  across the lattice, for both excitators and inhibitors. 

We now specialize on the stochastic, finite size dynamics, and hence assume $V_1$ to be finite. When $D=0$  the stochastic trajectories on each node are formally disentangled. 
Excitators (reps. inhibitors) execute almost regular oscillations about their deterministic equilibrium. These oscillations are termed in the literature quasi-cycles \cite{mckanenewman,dauxois} and follow a resonant amplification process triggered by the endogenous component of noise \cite{bartlett}. The amplitude of the oscillations scales as $1/\sqrt{V_1}$ and the associated frequency approximately reads $\omega_0 = r/4$, the imaginary part 
of the Jacobian eigenvalues in the uncoupled, $D=0$, setting. A remarkably different scenario is faced when turning the coupling active. We will in particular operate for $D<r/2$, the homogeneous fixed point being 
therefore stable. The degenerate component of the Jacobian spectrum returns an additional frequency $\omega_1 = \sqrt{\frac{r}{8}\left( \frac{r}{2}-D \right) }$ which can be continuously modulated, in the range $[0, \omega_0]$, as function of $D$. This observation is central to understand the emerging stochastic dynamics: the internal noise seeds in fact giant quasi-cycles, with tunable frequency and growing amplitude across the lattice. The system spontaneously behaves as an effective, stochastic driven pacemaker, a non trivial self-organized dynamics that we shall hereafter demonstrate. 

Under the  linear noise approximation (LNA) \cite{vankampen, gardiner}, stochastic effects act as linear deviations from the deterministic solution. Set $x_i=x^*+ \xi_i / \sqrt{V_i}$ 
and $y_i=y^*+ \eta_i / \sqrt{V_i}$, where $(\xi_i,\eta_i)$ stand for the stochastic perturbation. Inserting the 
above ansatz in (\ref{LE}) and performing an expansion at the first order in $1/ \sqrt{V_1}$  (see SI for details) returns 
a set of linear Langevin equations for the fluctuations amount. Denoting ${\boldsymbol \zeta}=(\xi_1,\eta_1, ... , \xi_\Omega, \eta_\Omega)$, one eventually obtains:
\begin{equation}
\label{LLE}
\frac{d}{d \tau} \zeta_i = \sum_{j=1}^{2 \Omega} J_{ij} \zeta_j + \rho_i
\end{equation}
where $\rho_i$ is Gaussian noise with zero mean and correlator  $\langle \rho_i (\tau) \rho_j (\tau') \rangle = {\mathcal B_{ij}} \delta (\tau-\tau')$; ${\boldsymbol {\mathcal B}}$ is the diffusion matrix of the associated Fokker-Planck equation and it is defined in the SI.  Let us denote $\tilde{\zeta_i}(\omega)$ the Fourier transform of $\zeta_i(t)$. Then, one gets $\tilde{\zeta}_i (\omega) = \sum^{2 \Omega}_{j=1} \Phi^{-1}_{ij}(\omega) \tilde{\rho}_j(\omega)$ where  $\Phi_{ij}=- J_{ij}-i \omega \delta_{ij}$. One can hence calculate the power spectrum of fluctuations on node $i$:
\begin{equation}
\label{PS}
P_{i} (\omega)= <\tilde{\zeta}_i(\omega)\tilde{\zeta}_i^*(\omega)> = \sum^{2 \Omega}_{l=1}\sum^{2 \Omega}_{m=1} \Phi^{-1}_{il} (\omega)\delta_{lm} \left( \Phi^{\dag} \right)^{-1}_{mi}(\omega)
\end{equation}

In the first panel of Figure  \ref{fig2} the (normalized) power spectrum of excitators fluctuations on different nodes is plotted. Symbols refer to the numerical integration of equations (\ref{LE}) \cite{euler_maruyama}, while the solid lines follows the theoretical estimate (\ref{PS}).  The power spectrum on the first node (circles, black online) is centered in $\omega_0$ (rightmost vertical dashed line). The power spectrum on the second node (squares, red online) displays a bimodal profile. A second peak emerges in correspondence of $\omega_1$, leftmost vertical dashed line. Moving along the chain (pluses and diamonds), the bump in $\omega_0$ fades away, while the peak in $\omega_1$ gains in potency and gets progressively more localized. Individual trajectories as obtained on different nodes are superposed in the lower panel of Figure  \ref{fig2}: the amplification can be clearly appreciated by eye inspection. A movie is also annexed as SI to better visualize the displayed amplification process. Under the linear noise approximation, the maximum of the power spectrum diverges exponentially (not shown) along the chain. At the same time the width of the bell in $\omega_1$ becomes narrower and the profile converges asymptotically to a delta like distribution. Beatings and other spurious modulations are therefore progressively filtered,  as moving along the chain and building on the idealized linear approach:  the system is hence predicted to eventually behave as a veritable pacemaker. However, non linear terms do matter and eventually balance the growth, as  predicted within linear scenario. Indeed, the process of amplification is expected to come to an halt when the oscillations get large enough so as to feel the boundary at $x_i \simeq 0$ (resp. $y_i \simeq 0$). 
 
\begin{figure}
 \centering
 %\subfigure[]
   {\includegraphics[width=6.5cm]{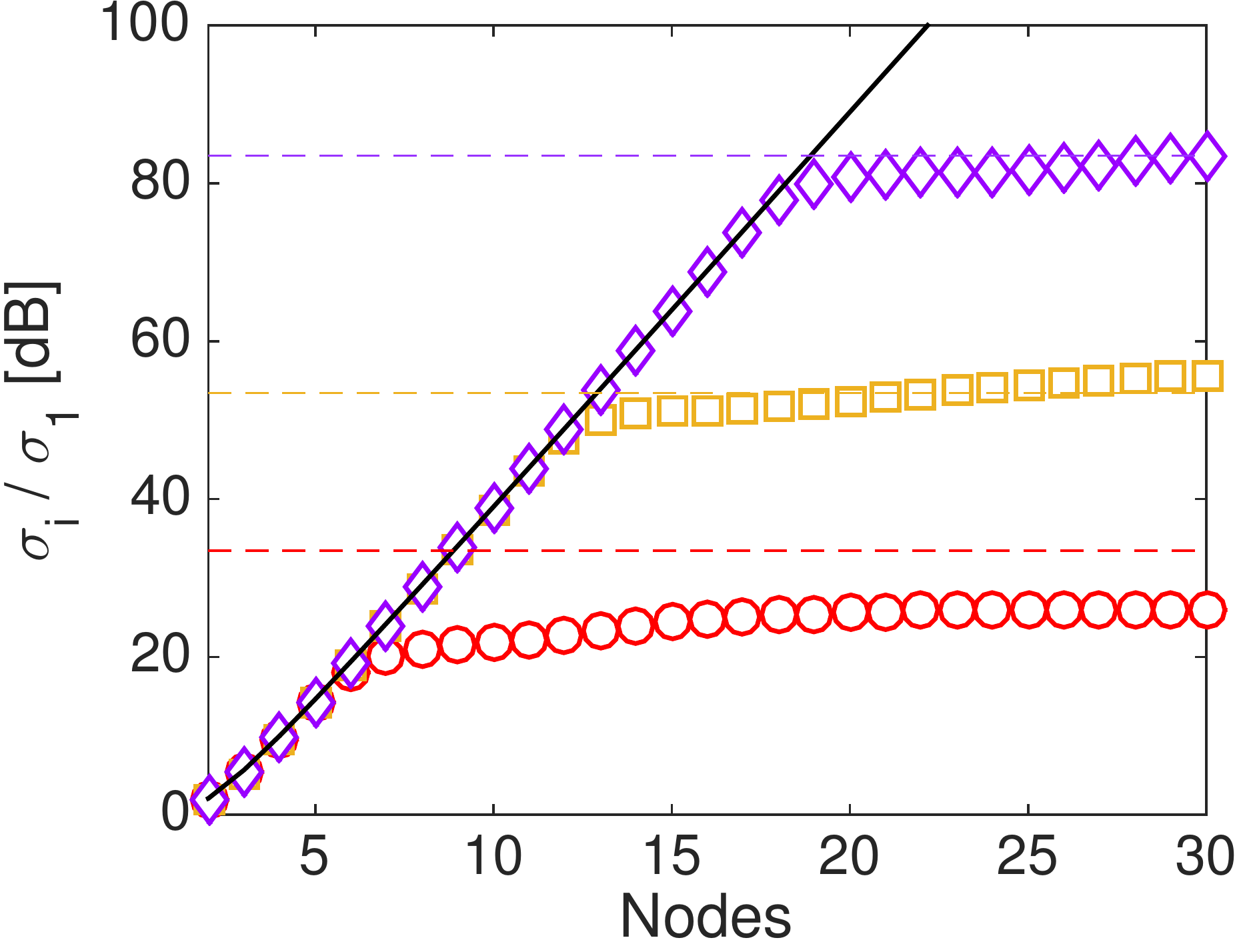}}
 %\hspace{5mm}
  %\centering
 %\subfigure[]
   {\includegraphics[width=6.5cm]{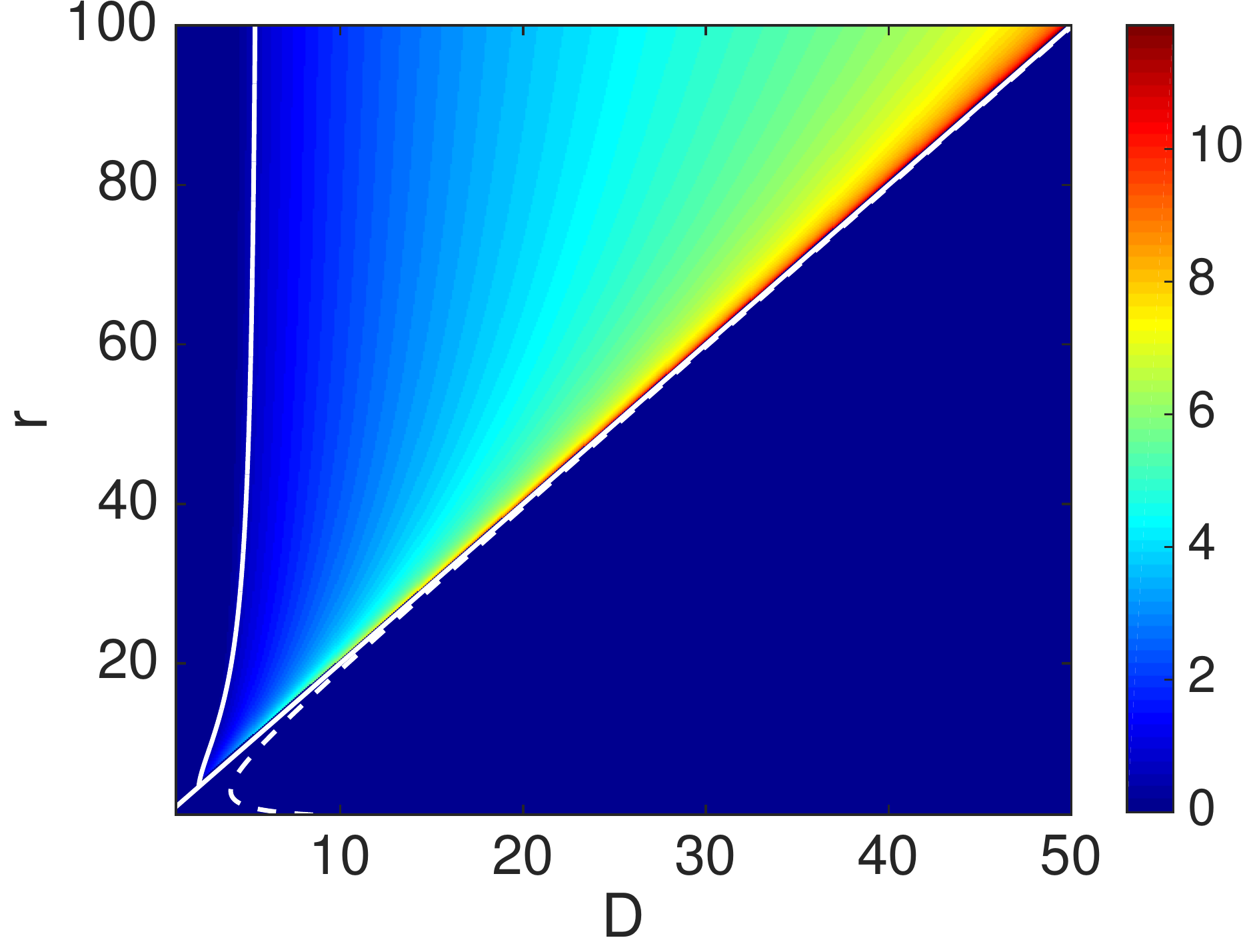}}
 %\hspace{5mm} 
  %\hspace{5mm}
   \caption{Upper panel: $\sigma_i / \sigma_1$ (in decibel [dB] logarithmic scale) is plotted against the index which identified the ordering of the nodes  along the lattice. Data refer to the exicitatory species. The solid line stands for the analytical estimate obtained under linear noise approximation. The amplification process is clearly exponential.  Symbols refer to direct integration of equations (\ref{LE}), for different choices of the volume $V_1$ ($10^6$, circles; $10^{12}$, squares; $10^{18}$, diamonds). The horizontal dashed lines show where the linear estimate predicts $\sigma_i \simeq 1/2$, namely when saturation is theoretically expected to occur. Here $D=10$, $r=50$. Lower panel: the rate of exponential amplifications (for the excitators) is  depicted in the plane $(r,D)$. The domain where the amplification is expected to take place are delimited by the two solid curves. The dashed line refers to $D_c$ vs. $r$. }
   \label{fig3}
  \end{figure}

To shed light on to this mechanism and quantify the amplification grade under the linear noise approximation, we set to consider the distribution of fluctuations $\Pi({\boldsymbol \zeta},t)$  around the deterministic equilibrium. As it is shown in the SI, $\Pi({\boldsymbol \zeta},t)$ obeys to a Fokker-Planck equation which can be self-consistently derived, via the van Kampen system size expansion. The solution of the Fokker-Planck equation  is a multivariate Gaussian that we can univocally characterize in terms of the associated first and second moments. It is immediate to show that the first moment converges to zero. We focus instead on the 
the $2 \Omega \times 2 \Omega$ family of second moments, defined as $\langle \zeta_l \zeta_m \rangle = \int \zeta_l \zeta_m \Pi d {\boldsymbol \zeta}$. A straightforward calculation (see SI) yields:

\begin{eqnarray}
\label{moments}
\frac{d}{d\tau}<\zeta^2_l>&=&+2<\zeta_l(J{\boldsymbol \zeta})_l>+{\mathcal B}_{ll} \\
\frac{d}{d\tau}<\zeta_l\zeta_m>&=&<\zeta_l(J{\boldsymbol \zeta})_m>+<\zeta_m(J{\boldsymbol \zeta})_l> \nonumber
\end{eqnarray}

for respectively the diagonal and off-diagonal ($l \ne m$) moments. The stationary values of the moments can be analytically computed by setting to zero the time derivatives on the left hand side of equations (\ref{moments}) and solving the linear system that is consequently obtained. We are in particular interested in accessing $\sigma_i=\sqrt{\langle \zeta_i^2 \rangle}$, the standard deviation of the fluctuations as displayed, around the deterministic equilibrium, on node $i$. The value of  $\sigma_i$, normalized to $\sigma_1$ and expressed in decibel [dB], is plotted against the nodes index along the lattice in Figure  \ref{fig3}, upper panel. The data refer to the excitatory species. The solid line stands for the analytical estimate, that implements the above strategy. Remarkably, the standard deviation of the fluctuations grows exponentially along the chain. Symbols refer instead to direct integration of equations (\ref{LE}), for different choices of the volume $V_1$. The agreement with the theory prediction based on the linear ansatz is excellent over a finite portion of the chain. When $\sigma_i \simeq 1/2$ (horizontal dashed line) the system senses the boundary, non linearities come into play and induce the observed saturation. By increasing $V_1$, one reduces the amplitude of the endogenous fluctuations: the signal has therefore to travel through a larger set of contiguous nodes before the amplitude of the oscillation  can hit the extinction edge. As a consequence, the linear approximation holds over a larger portion of the scrutinized chain.
The rate of exponential growth (relative to the excitators species), as predicted by the linear theory, is plotted with an appropriate color code, in the reference parameters plane $(r,D)$, see lower panel of Figure \ref{fig3}. The amplification takes place within a bounded region in $(r,D)$, as delimited by the two solid (white) lines. The straight line that sets the rightmost frontier of the amplification domain is  obtained as $r=D/2$, namely the condition of existence of a complex imaginary part in the degenerate eigenvalues $\lambda_{3,4}$ (which in turn select the frequency $\omega_1$ to be amplified). The boundary that delimits the region of interest on the left follows a closed analytical estimate, obtained by truncating long range correlations in the estimate of the multivariate moments to nearest neighbors (details in the SI). The dashed (white) line refers to $D_c$ vs. $r$ and it is depicted for the sake of completeness. Similar results (not shown) apply to the inhibitors. 

The noise assisted amplification process that we have here characterized is very flexible and can be configured in different schemes.  By augmenting the volumes of the nodes along the chain, and so consequently tuning the additional control parameters $\gamma_i$, we can amplify  virtually any harmonic of $\omega_0$ (or, alternatively, $\omega_1$).  This possibility is demonstrated in the SI. The amplification pattern can also take place on a frequency comb, by appropriately assigning the relative weight $\gamma_i$, see SI.  Following a similar strategy, it is also possible to focus the amplification on frequencies larger than $\omega_0$.

In all inspected cases, the self-sustained amplification is  fueled by the inherent component of noise, stemming from finite size corrections. At variance, one could imagine to assemble a device that operate in the deterministic $V_i \rightarrow \infty$  limit. If $D<D_c$, the system is frozen in its homogeneous equilibrium, the concentration of both $x_i$ and $y_i$ being identical to $1/2$ on each node. Assume now that a 
perturbation, limited in time and modest in amplitude, hit on the first node. For demonstrative purposes we exemplify the perturbation as a noisy signal, drawn from a  random uniform distribution. The disturbance propagate along the chain and gets magnified, as follows the scheme that we outlined above, exciting on site oscillations at a given frequency $\omega_1$, that could be freely tuned by acting e.g. on $D$. Such an apparatus could efficaciously act as a signal detector, a possibility that is demonstrated in the movie annexed as SI. Even more interesting, one could foresee the possibility of assembling a detector that exploits parallel lines of detection. On each line a different value of the coupling $D$ could be enforced. In doing so, from the trace of the amplified signal at the end of the chain (processed with a standard frequency analyzer), it could be possible to identify the node (hence the chain) where the perturbation hit. This observation opens up the perspective to define a novel class of detectors that could spatially resolve low intensity alerts.

As a final point, we will elaborate on a consistent thermodynamic interpretation of the process that underlies the spontaneous generation of giant quasi-oscillations. Our analysis follows the approach pioneered by \cite{thermo0,thermo, thermo1} to study the thermodynamics of far-from-equilibrium systems,  which are microscopically amenable to stochastic continuous time Markovian processes. Given the probability density $P({\boldsymbol v}, \tau)$ that satisfies the Fokker-Planck equation [Eq. (1) in the SI], we define the entropy $S(\tau)=-\int P({\boldsymbol v},\tau)\ln P({\boldsymbol v},\tau) d{\boldsymbol v}$.  A straightforward manipulation yields $dS/dt=\Pi_S -\Phi_S$, where (i) $\Pi_S$ is positive defined and represents the rate of entropy production due to the non-conservative forces at play and (ii) $\Phi_S$ stands for the entropy flux, which is positive if the entropy flows from the system to the environment. Explicit expressions for both $\Pi_S$ and $\Phi_S$ are available, as reported in the SI. A stationary balance is  attained when $\Phi_S=\Pi_S$, a condition that proves equivalent to imposing $\sum \frac{\partial}{\partial v_i} I_i = 0$, where 
$I_i = A_i P - \frac{1}{2} B_{ii} \frac{\partial}{\partial v_i} P$ is the probability density current associated to the Fokker-Planck equation [Eq. (1) in the SI]. The condition of solenoidal  current, ${\boldsymbol \nabla} \cdot {\boldsymbol I}=0$ is indeed met when the Fokker-Planck equation attains its non trivial dynamical equilibrium ($I_i \ne 0$). In other words, the observed amplification stems from a genuine noise driven out-of-equilibrium process, the neuromorphic device working  under stationary operating conditions. The rate of entropy production as computed analytically under a linear prescription grows exponentially, see SI. A cross-over towards a non exponential regime is eventually observed when non linearities become prominent, in complete agreement with the insight gained under a purely dynamical angle.

In conclusion, we have here shown how a minimal model of neuronal population dynamics can be assembled to result in a fully tunable amplifier. The device extracts energy from the finite size bath and operates as an out of equilibrium thermal machine. A spatially distributed detector of low intensity noisy signals can be foreseen which exploits the same, neuromorphic inspired, architecture. Extension to settings where the network of couplings results 
in a  non-normal adjacency matrix \cite{non_norm1, non_norm2} are currently under investigations.

\section*{Acknowledgments}
The authors acknowledge financial support from H2020-MSCA-ITN-2015 project COSMOS  642563.

\appendix

\section{On the Kramers-Moyal approximation.}

The master equation that governs the evolution of $P({\boldsymbol v},t)$ can be written as:
\[
\frac{\partial P}{\partial t}=(\Gamma_1+\Gamma_2)P
\]
where the operators $\Gamma_1$ and $\Gamma_2$ are given by
\[
\Gamma_1=\sum_{i=1}^{\Omega} (\epsilon_{{x_i}}^{+}-1)T({x_i}-\frac{1}{V_i}|{\boldsymbol v})+(\epsilon_{{x_i}}^{-}-1)T({x_i}+\frac{1}{V_i}|{\boldsymbol v})
\]
\[
\Gamma_2=\sum_{i=1}^{\Omega} (\epsilon_{{y_i}}^{+}-1)T({y_i}-\frac{1}{V_i}|{\boldsymbol v})+(\epsilon_{{y_i}}^{-}-1)T({y_i}+\frac{1}{V_i}|{\boldsymbol v})
\]
and the step operators are defined as 
\begin{eqnarray*}
\epsilon_{x_i}^\pm f({x_i},{y_i}) &=& f({x_i} \pm \frac{1}{V_i},{y_i}) \\
\epsilon_{y_i}^\pm f({x_i},{y_i}) &=& f({x_i},y_i \pm \frac{1}{V_i}) 
\end{eqnarray*}

For large enough $V_i$, one can approximate the step operators as:
 \begin{eqnarray*}
\epsilon_{{x_i}}^\pm &\approx& 1 \pm \frac{1}{V_i}\frac{\partial}{\partial {x}_i}+\frac{1}{2V_i^2} \frac{\partial^2}{\partial {x}_i^2} \\
\epsilon_{{y_i}}^\pm &\approx& 1 \pm \frac{1}{V_i}\frac{\partial}{\partial {y}_i}+\frac{1}{2V_i^2} \frac{\partial^2}{\partial {y}_i^2}
\end{eqnarray*}

By inserting the above expressions into the master equation, performing the calculations and introducing $\tau=\frac{t}{V_1}$ yields the following Fokker-Planck equation:

\begin{equation}
\label{FP}
\frac{\partial P}{\partial \tau}= -\sum_{i=1}^{2\Omega} \frac{\partial}{\partial {v}_i} A_i P +\sum_{i,j=1}^{2\Omega}\frac{1}{2V_1}\frac{\partial^2}{\partial {v}_i\partial {v}_j} B_{ij}P
\end{equation}

where
\[
{\boldsymbol A}=
\begin{pmatrix}
... \\
...  \\
\frac{1}{\gamma_i} \big(T({x_i}+\frac{1}{V_i}|{\boldsymbol v})-T({x_i}-\frac{1}{V_i}|{\boldsymbol v})  \big) \\
\frac{1}{\gamma_i}\big(T({y_i}+\frac{1}{V_i}|{\boldsymbol v})-T({y_i}-\frac{1}{V_i}|{\boldsymbol v})  \big) \\
... \\
... \\
\end{pmatrix}
\]
and ${\boldsymbol B}$ is a diagonal $2 \Omega \times 2 \Omega$ matrix made of  $\Omega$ distinct $2 \times 2$ blocks ${\boldsymbol B}_i$ with entries:

\begin{eqnarray*}
({\boldsymbol B}_i)_{11}&=&\frac{1}{\gamma_i^2}\big(T({x_i}+\frac{1}{V_i}|{\boldsymbol v})+T({x_i}-\frac{1}{V_i}|{\boldsymbol v})  \big) \\
({\boldsymbol B}_i)_{12}&=&({\boldsymbol B}_i)_{21}=0 \\
({\boldsymbol B}_i)_{22}&=& \frac{1}{\gamma_i^2}\big(T({y_i}+\frac{1}{V_i}|{\boldsymbol v})+T({y_i}-\frac{1}{V_i}|{\boldsymbol v})  \big) 
\end{eqnarray*}
for i=$1, ..., \Omega$.

By making use of the explicit form of the transition rates:

\begin{eqnarray*}
T({x_i}+\frac{1}{V_i}|{\boldsymbol v}) &=& f(s_{x_i}) \\
T({x_i}-\frac{1}{V_i}|{\boldsymbol v})  &=& x_i \\
T({y_i}+\frac{1}{V_i}|{\boldsymbol v}) &=& f(s_{y_i}) \\
T({y_i}-\frac{1}{V_i}|{\boldsymbol v})  &=& y_i 
\end{eqnarray*}

one readily obtains, from the above Fokker-Planck equation, the equivalent Langevin equations as reported in the main body of the paper.

\section{Stability of the homogeneous mean field solution.}

In the thermodynamic limit $V_1 \rightarrow \infty$, the examined stochastic system reduces to the following deterministic system: 

\begin{eqnarray*}
\frac{d}{d\tau}{x}_i&=&\frac{1}{\gamma_i}\big[f(s_{x_i})-{x}_i  \big] \\
\frac{d}{d\tau}{y}_i&=&\frac{1}{\gamma_i}\big[f(s_{y_i})-{y}_i  \big]\\ 
\end{eqnarray*}

which, as stated in the main body of the paper, admits $x_i =y_i = 1/2$ $\forall i$, as an homogenous fixed point. To determine the stability of the system we carry out the linear stability analysis and obtain 

\begin{equation}
{\boldsymbol J}=
\begin{pmatrix}
{\boldsymbol E_1}  & 0& 0 & 0 & 0  \\
{\boldsymbol S_2} & {\boldsymbol E_2} & 0& 0 & 0  \\
0 & {\boldsymbol S_3} & {\boldsymbol E_3} & 0 & 0   \\
0 & 0 & \ddots & \ddots & 0 \\
0 & 0 & 0 & {\boldsymbol S_\Omega} &  {\boldsymbol E_\Omega} \\
\end{pmatrix}
\end{equation}

where

\[
{\boldsymbol E_1} =
\begin{pmatrix}
-1 & -\frac{r}{4} \\
\frac{r}{4} & -1 
\end{pmatrix}
\]

\[
{\boldsymbol E_i} =
\begin{pmatrix}
-\frac{1+D/4}{\gamma_i} & -\frac{r-D}{4\gamma_i} \\
\frac{r-D}{4\gamma_i} & -\frac{1-D/4}{\gamma_i} 
\end{pmatrix}
\]

\[
{\boldsymbol S_i} =\frac{D}{4\sqrt{\gamma_i\gamma_{i-1}}}
\begin{pmatrix}
1& -1 \\
1& -1 
\end{pmatrix}
\]

The characteristic polynomial of ${\boldsymbol J}$ writes 
$0=det(J-\lambda I)=det({E_1}-\lambda I)\prod_{i=2}^{\Omega} det(E_i-\lambda I)$. 
The first term in the preceding expression gives a quadratic equation for $\lambda$, 
$(\lambda+1)^2+\frac{r^2}{16}=0$. This latter yields $\lambda_{1,2}=-1\pm i\frac{r}{4}\equiv-1\pm i\omega_0$. 
The remaining eigenvalues are obtained by solving the following $\Omega$ equations:
$$\big( \frac{1+D/4}{\gamma_i}+\lambda \big)\big( \frac{1-D/4}{\gamma_i}+\lambda \big)+\frac{(r-D)^2}{16\gamma_i^2}=0$$
allowing one to immediately obtain: 
$$(\lambda_i)_{3,4}=\frac{1}{\gamma_i}\left[-1\pm \sqrt{-\frac{r}{8}(\frac{r}{2}-D)}\right].$$ 
Notice that for $V_i=V_1$ $\forall i$ (or, equivalently, $\gamma_i=1$, $\forall i$), $(\lambda_i)_{3,4} \equiv \lambda_{3,4}$, 
as stated in the main body of the paper. By additionally requiring $D<r/2$, $\lambda_{3,4}=-1 \pm i \omega_1$, with 
$\omega_1=\sqrt{\frac{r}{8}\left( \frac{r}{2}-D \right) }$.

\subsection{Linear noise approximation (LNA)}

To perform the linear noise approximation we set:

\begin{eqnarray}
x_i&=&x^*+\frac{\xi_i}{\sqrt{V_i}}=x^*+\frac{1}{\sqrt{V_1}}\frac{\xi_i}{\sqrt{\gamma_i}}\\
y_i&=&y^*+\frac{\eta_i}{\sqrt{V_i}}=y^*+\frac{1}{\sqrt{V_1}}\frac{\eta_i}{\sqrt{\gamma_i}}
\end{eqnarray}

Insert the above ansatz in the non linear function $f(\cdot)$ and performing the expansion at the first order in $1/\sqrt{V_1}$ yields:

\begin{eqnarray*}
f(s_{x_i})&\approx& \frac{1}{2}+\frac{1}{4 \sqrt{V_1}} \left[-\frac{r\eta_i}{\sqrt{\gamma_i}}+D  \Delta \xi_i -D  \Delta \eta_i \right]+  \mathcal{O}(\frac{1}{V_1}) \\
f(s_{x_i})&\approx& \frac{1}{2}+\frac{1}{4 \sqrt{V_1}} \left[\frac{r\xi_i}{\sqrt{\gamma_i}}+D  \Delta \xi_i -D  \Delta \eta_i \right] +   \mathcal{O}(\frac{1}{V_1}) \\
\end{eqnarray*}

where $\Delta \xi_i \equiv (\frac{\xi_{i-1}}{\sqrt{\gamma_{i-1}}}-\frac{\xi_{i}}{\sqrt{\gamma_i}})$ and $\Delta \eta_i \equiv (\frac{\eta_{i-1}}{\sqrt{\gamma_{i-1}}}-\frac{\eta_{i}}{\sqrt{\gamma_i}})$. Here,
use has been made of the condition $f(0)=1/2$ and $f'(0)=f(0)(1-f(0))=1/4$, $f'(\cdot)$ labelling the derivative of $f(\cdot)$.
The nonlinear amplitudes that characterize the multiplicative noise in the Langevin equations (see main paper) reduces to constant factors under the linear noise approximation. Specifically: 

\begin{eqnarray*}
\frac{1}{\gamma_i\sqrt{V_1}}\sqrt{{x}_i+f(s_{X_i})}\approx \frac{1}{\gamma_i\sqrt{V_1}}+\mathcal{O}(\frac{1}{V_1})\\
\frac{1}{\gamma_i\sqrt{V_1}}\sqrt{{y}_i+f(s_{Y_i})}\approx \frac{1}{\gamma_i\sqrt{V_1}}+\mathcal{O}(\frac{1}{V_1})
\end{eqnarray*}

Building on the above one can write the linear version of the non linear Langevin equations. To this end, introduce 
${\boldsymbol \zeta}=(\xi_1,\eta_1, ... , \xi_\Omega, \eta_\Omega)$ and write:  

\begin{equation}
\label{LLE}
\frac{d}{d \tau} \zeta_i = \sum_{j=1}^{2 \Omega} J_{ij} \zeta_j + \rho_i
\end{equation}
 
 where $J_{ij}$ are the entries of the Jacobian matrix ${\boldsymbol J}$ and  $\rho_i$ is Gaussian noise with zero mean and correlator  
 $\langle \rho_i (\tau) \rho_j (\tau') \rangle = {\mathcal B_{ij}} \delta (\tau-\tau')$.  ${\mathcal B_{ij}}$ are the entries of the diffusion matrix ${\boldsymbol {\mathcal B}}$ defined as:
 
  \[
{\boldsymbol {\mathcal B}}=
\begin{pmatrix}
{\boldsymbol I}_{2\times2} & 0& 0 & 0 & 0  \\
0 & \frac{1}{\gamma_2}{\boldsymbol I}_{2\times2}  & 0& 0 & 0  \\
0 & 0 & \frac{1}{\gamma_3}{\boldsymbol I}_{2\times2} & 0 & 0   \\
0 & 0 & \ddots & \ddots & 0 \\
0 & 0 & 0 & 0 &  \frac{1}{\gamma_\Omega}{\boldsymbol I}_{2\times2} \\
\end{pmatrix}
\]

where ${\boldsymbol I}_{2\times2}$ stands for the $2 \times 2$ identity matrix.  The linear Langevin equations (\ref{LLE}) are equivalent to the following Fokker-Planck for the probability distribution
function $\Pi({\boldsymbol \zeta},\tau)$ of fluctuations:

\begin{equation}\label{eq:Fokker_Planck_with_different_volumes}
\frac{\partial}{\partial\tau}\Pi=-\sum_{i=1}^{2\Omega} \frac{\partial}{\partial\zeta_i}\big[(J\zeta)_i\Pi \big]+\frac{1}{2}\sum_{i,j=1}^{2\Omega}\frac{\partial^2}{\partial\zeta_i\partial\zeta_j}{\mathcal B}_{ij}\Pi
\end{equation}

It is worth emphasising that the above equation could be also derived by performing a van Kampen expansion of original master equation for the probability $P({\boldsymbol v},t)$.

\subsection{Computing the moments of the Gaussian multivariate distribution $\Pi$}

We shall here derive the dynamical equations that control the evolution of the moments of the distribution $\Pi$. Focus on the first moment, by
multiplying  equation (\ref{eq:Fokker_Planck_with_different_volumes}) by $\zeta_k$ and integrating over ${\boldsymbol \zeta}$.
The left hand side  of  equation yields:
\[
\int d{\boldsymbol \zeta}\zeta_k \frac{\partial}{\partial\tau}\Pi=\int d{\boldsymbol \zeta}\frac{\partial}{\partial\tau}\Pi\zeta_k=\frac{d}{d\tau}\int d{\boldsymbol \zeta}\zeta_k\Pi=\frac{d}{d\tau}<\zeta_k>
\]
The right hand side can be split into two parts.  Under mild assumptions for $\Pi$, the drift term returns:
\[
-\sum_{i=1}^{2\Omega} \int d{\boldsymbol \zeta} \zeta_k\frac{\partial}{\partial\zeta_i}\big[ (J{\boldsymbol \zeta})_i\Pi\big]
\]
The contribution $i=k$ amounts to:
\[
\begin{split}
\int &d{\boldsymbol \zeta} \zeta_k\frac{\partial}{\partial\zeta_k}\big[ (J{\boldsymbol \zeta})_k\Pi\big]=\\
&=\int \prod_{j \neq k} d\zeta_j \int d\zeta_k \zeta_k\frac{\partial}{\partial\zeta_k}\big[ (J{\boldsymbol \zeta})_k\Pi\big]=\\
&=-\int \prod_{j \neq k} d\zeta_j \int d\zeta_k\big[ (J{\boldsymbol \zeta})_k\Pi\big]= \\
&=-\int d{\boldsymbol \zeta}\big[ (J{\boldsymbol \zeta})_k\Pi\big]=-<(J{\boldsymbol \zeta})_k>
\end{split}
\]
while the terms with $i\neq k$ give no contributions. In fact:
\[
\int \prod_{j \neq k,i} d\zeta_j \int d\zeta_k \zeta_k \int d\zeta_i \frac{\partial}{\partial\zeta_i}\big[ (J{\boldsymbol \zeta})_i\Pi\big]=0
\]
It is then straightforward to conclude that the diffusion terms returns no contributions, because $\Pi$ decays fast enough at the boundaries. 
 Summing up, we therefore obtain the linear equations:
\[
\frac{d}{d\tau}<\zeta_k>=<(J{\boldsymbol \zeta})_k>=\sum_{j=1}^{2\Omega}J_{kj}<\zeta_j>
\]
The unique stationary (stable) solution is therefore  $<\zeta_k>=0 \quad \forall k$.

An identical procedure can be followed to evaluate the second moments of the distribution, namely $<\zeta_l\zeta_m>$. To this end we multiply equation (\ref{eq:Fokker_Planck_with_different_volumes}) by $\zeta_l\zeta_m$ and integrate over ${\boldsymbol \zeta}$. 
In analogy with the above, the left hand side of the equation returns: 
\[
\int d{\boldsymbol \zeta} \zeta_l\zeta_m \frac{\partial}{\partial\tau}\Pi= \frac{d}{d\tau}<\zeta_l\zeta_m>
\]
When it comes to the drift term, we shall focus first on the diagonal, $l=m$, contributions:

\[
-\sum_{i=1}^{2\Omega} \int d{\boldsymbol \zeta} \zeta_l^2 \frac{\partial}{\partial\zeta_i}\big[ (J{\boldsymbol \zeta})_i\Pi \big]
\]   
For $i=l$, we get:
 \[
 \begin{split}
 \int &d{\boldsymbol \zeta} \zeta_l^2 \frac{\partial}{\partial\zeta_l}\big[ (J{\boldsymbol \zeta})_l\Pi \big]=\int \prod_{j \neq l} d\zeta_j \int d\zeta_l \zeta_l^2 \frac{\partial}{\partial\zeta_l}\big[ (J{\boldsymbol \zeta})_l\Pi \big]=\\
 &=-2\int \prod_{j \neq l} d\zeta_j \int d\zeta_l \zeta_l (J{\boldsymbol \zeta})_l \Pi=-2<\zeta_l(J{\boldsymbol \zeta})_l>
\end{split} 
 \]
while for $i\neq l$ one finds:
 \[
 \begin{split}
 \int &d{\boldsymbol \zeta} \zeta_l^2 \frac{\partial}{\partial\zeta_i}\big[ (J{\boldsymbol \zeta})_i\Pi \big] = \\
 &=\int \prod_{j \neq l,i} d\zeta_j \int d\zeta_l \zeta_l^2\int d\zeta_i \frac{\partial}{\partial\zeta_i}\big[ (J{\boldsymbol \zeta})_i\Pi \big]=0
\end{split} 
 \]
 Consider now the contribution of the drift  to the off diagonal elements ($l\neq m$), namely:
 \[
 -\sum_{i=1}^{2\Omega} \int d{\boldsymbol \zeta} \zeta_l \zeta_m \frac{\partial}{\partial\zeta_i}\big[ (J{\boldsymbol \zeta})_i\Pi \big]
 \]
 For $i=l$, one gets:
 \[
 \begin{split}
 \int & \prod_{j \neq l,m} d\zeta_j \int d\zeta_m \zeta_m \int d\zeta_l  \zeta_l \frac{\partial}{\partial\zeta_l}\big[ (J{\boldsymbol \zeta})_l\Pi \big]=\\
 &=- \int \prod_{j \neq l,m} d\zeta_j \int d\zeta_m \zeta_m \int d\zeta_l \big[ (J{\boldsymbol \zeta})_l\Pi \big]=\\
& =-<\zeta_m (J{\boldsymbol \zeta})_l>
\end{split} 
 \]
 The other case of interest, $i=m$, is easy to treat, as it amounts to swapping $l$ and $m$. Finally, for $i\neq l,m$ the drift term returns a null contribution:
 \[
 \int \prod_{j \neq m,l,i} d\zeta_j \int d\zeta_m \zeta_m \int d\zeta_l  \zeta_l \int d \zeta_i \frac{\partial}{\partial\zeta_i}\big[ (J{\boldsymbol \zeta})_i\Pi \big]=0
 \]

 Let us now turn to considering the contribution of the diffusion terms in the Fokker-Planck equation. Since ${\boldsymbol {\mathcal B}}$ is diagonal,  a non trivial contribution is solely found for $l=m$:
 
\[
\frac{1}{2} \sum_{i=1}^{2\Omega} \int d{\boldsymbol \zeta} \zeta_l^2 \frac{\partial^2}{\partial\zeta_i^2}{\mathcal B}_{ii}\Pi
\]  
For $i=l$, we have:
\[
\begin{split}
\frac{1}{2} \int &\prod_{j \neq l} d\zeta_j \int d\zeta_l \zeta_l^2 \frac{\partial^2}{\partial\zeta_l^2}{\mathcal B}_{ll}\Pi =\\
&=-2\frac{1}{2}\int \prod_{j \neq l} d\zeta_j \int d\zeta_l \zeta_l \frac{\partial}{\partial\zeta_l}{\mathcal B}_{ll}\Pi= \\
&=\int \prod_{j \neq l} d\zeta_j \int d\zeta_l {\mathcal B}_{ll}\Pi={\mathcal B}_{ll}\int d{\boldsymbol \zeta}\Pi={\mathcal B}_{ll}
\end{split}
\]
where use has been made of the condition of normalization for the distribution $\Pi$. The case $i\neq l$ yields no contribution as:
\[
\frac{1}{2} \int \prod_{j \neq l,i} d\zeta_j \int d\zeta_l \zeta_l^2 \int d\zeta_i \frac{\partial^2}{\partial\zeta_i^2}{\mathcal B}_{ii}\Pi=0
\]
Collecting all terms together we end up with the equations for the second moments reported in the main body of the paper.

\subsection{Analytical estimate for the leftmost boundary of the amplification domain.}

Computing the moments of the multivariate Gaussian that characterizes the stationary distribution of fluctuations under the linear noise approximation, imply solving a $2 \Omega  \times 2 \Omega$ problem. To gain analytical insight into the problem (with reference to the setting $\gamma_i=1$), one can operate a drastic simplification by solely accounting for nearest neighbors correlations. In doing so, one obtains a $7 \times 7$ linear system, which we do not write here explicitly because it involves lengthy expressions. Due to the structure of the problem, the $7 \times 7$ system rigorously  reduces to an effective map, from a given node to the next one, for the reference quantities ${\boldsymbol w}_i = (<\xi_i^2>,<\eta_i^2>,<\xi_i \eta_i>)$. More concretely, one can recast the problem in the form:
\[
{\boldsymbol w}_{i+1}={\boldsymbol A} {\boldsymbol w}_i+{\boldsymbol r}
\]
where ${\boldsymbol A}$ (not given here explicitly) is non diagonalizable, it has rank $2$ and eigenvalues $0,\lambda$, with:
\[
\lambda=\frac{ - 2\, D^3\, r^3 + D^2\, r^4 + 80\, D^2\, r^2}{128\, D^2\, r^2 - 128\, D\, r^3 - 2048\, D\, r + 32\, r^4 + 1024\, r^2 + 8192}
\]
To solve the problem one can reduce ${\boldsymbol A}$ to a Jordan normal form ${\boldsymbol {\mathcal A}}$. It can be in fact shown that a matrix ${\boldsymbol {P}}$ exists such that ${\boldsymbol {\mathcal A}}={\boldsymbol {P}}^{-1}{\boldsymbol {A}}{\boldsymbol {P}}$ 

By operating the change of variables ${\boldsymbol {q}}_i={\boldsymbol {P}}^{-1}{\boldsymbol {w}}_i$ and defining 
${\boldsymbol {R}}={\boldsymbol {P}}^{-1}{\boldsymbol {r}}$ one gets: 
\[
{\boldsymbol q}_{i+1}={\boldsymbol {\mathcal A}} {\boldsymbol q}_i+{\boldsymbol R}
\]
that it can be shown to yield:
\[
\begin{cases}
q_{i+1}^{(1)}=q_{i}^{(2)}+R^{(1)} \\
q_{i+1}^{(2)}=R^{(2)} \\
q_{i+1}^{(3)}=\lambda q_{i}^{(3)}+R^{(3)}
\end{cases}
\]

where ${\boldsymbol q}_i \equiv (q_i^{(1)}, q_i^{(2)},q_i^{(3)})$  and ${\boldsymbol R} \equiv (R^{(1)}, R^{(2)},R^{(3)})$.
Solving the above system and going back to the original variables, one eventually gets:

\[
\begin{split}
<\xi^2_i>&=P_{13}\big(q_3(0)+\frac{R^{(3)}}{\lambda-1}\big)\lambda^i+\\
&+P_{11}(R^{(1)}+R^{(2)})+P_{12}R^{(2)}-P_{13}\frac{R^{(3)}}{\lambda-1}
\end{split}
\]
\[
\begin{split}
<\eta^2_i>&=P_{23}\big(q_3(0)+\frac{R^{(3)}}{\lambda-1}\big)\lambda^i+\\
&+P_{21}(R^{(1)}+R^{(2)})+P_{22}R^{(2)}-P_{23}\frac{R^{(3)}}{\lambda-1}
\end{split}
\]
\[
\begin{split}
<\eta_n\xi_i>&=P_{33}\big(q_3(0)+\frac{R^{(3)}}{\lambda-1}\big)\lambda^i+\\
&P_{31}(R^{(1)}+R^{(2)})+P_{32}R^{(2)}-P_{33}\frac{R^{(3)}}{\lambda-1}
\end{split}
\]

The amplification is hence lost if $|\lambda| \le 1$. The leftmost solid (white) line in Figure 3 in the main body of the paper (lower panel) corresponds to the limiting condition $\lambda=1$. The boundary of the domain where the amplification takes place is adequately reproduced, an observation that supports a posteriori the validity of the approximations involved in the analysis.

\subsection{Amplifying the harmonics of $\omega_0$.}

To amplify the harmonics of $\omega_0$ for any given $D$, within the domain deputed to the amplification, we can modulate the volumes of the nodes, following the strategy discussed below. Label $V_1$ the volume of the first node. Recall that $\omega_1=\sqrt{\frac{r}{8} \left(\frac{r}{2}-D\right)}$ identifies the frequency that gets amplified when the volumes are forced to be identical, or, equivalently, when $\gamma_i=1$ $\forall i$. To instigate the emergence of a second peak in $\omega_0/2$, on the second node of the lattice,  one needs to impose the condition $\frac{\omega_1}{\gamma_2} \equiv \frac{\omega_0}{2}$ which readily translates in $V_2=2V_1 \frac{\omega_1}{\omega_0}$. To enforce the amplification of a train of successive harmonics one can expand on the above recipe and eventually obtain the following condition for the relative modulation of the volumes:

\begin{equation}
\label{VHGain}
V_i = 2^{i-1} \frac{\omega_1}{\omega_0}V_1 \qquad i \ge 2
\end{equation}

\begin{figure}
 \centering
   {\includegraphics[width=7.5cm]{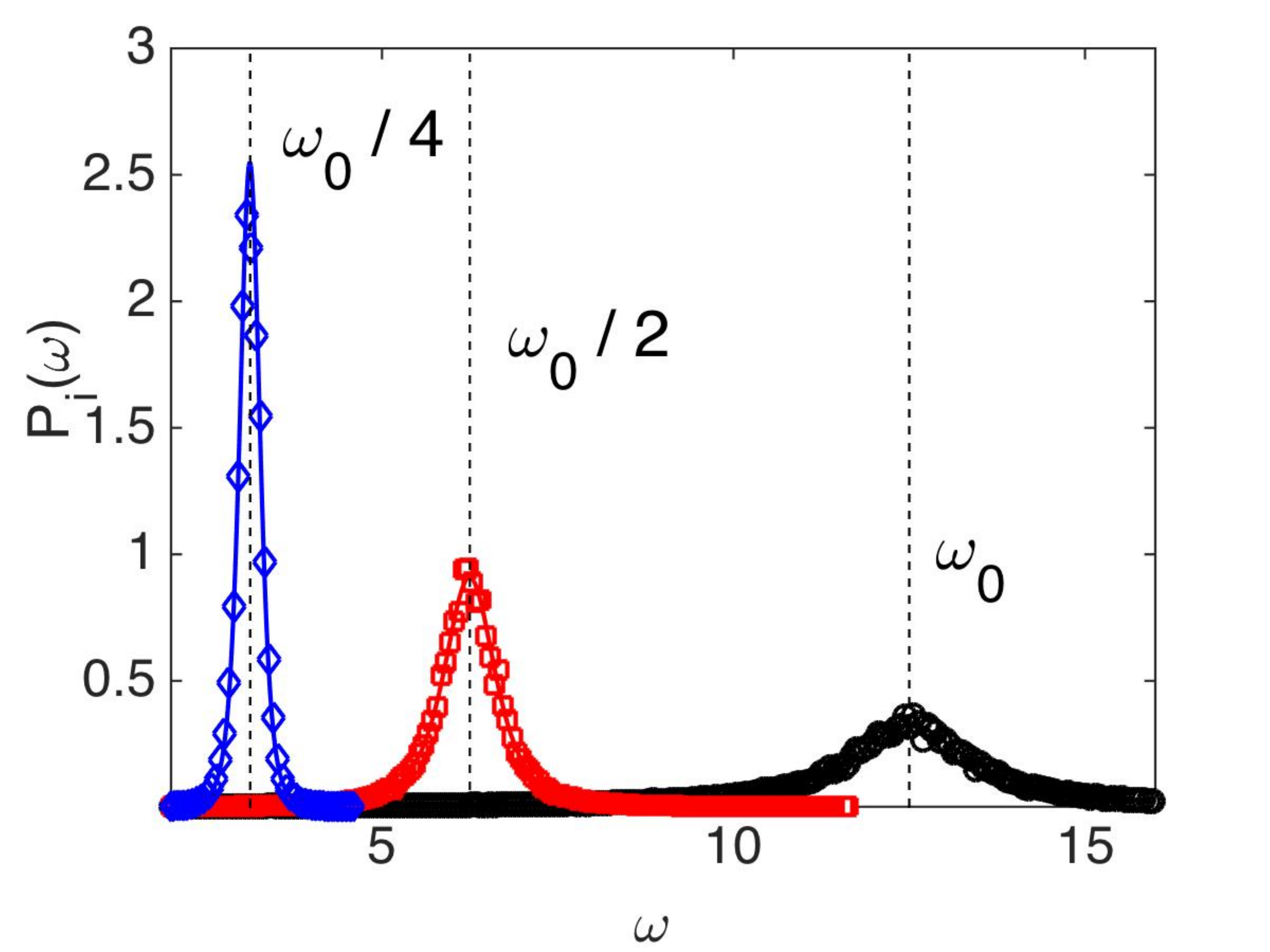}}
   \caption{Amplifying the harmonics of $\omega_0$, following the scheme that yields to equation (\ref{VHGain}). The power spectra of fluctuations on different nodes (see text) are displayed. Symbols refer to direct simulations and the solid lines to the theory prediction. 
}
\label{fig4}
\end{figure}

In practice, to allow for the amplification to produce significant intensities of the signal at each frequency, one could keep the volumes constant over a few consecutive nodes, before increasing the size of the volumes of the successive set of nodes, as prescribed by formula (\ref{VHGain}). In Figure  \ref{fig4} we assumed a sequence of nodes with volumes $(V_1, V_2, V_2, V_3, V_3, V_3)$. The power spectra depicted in Figure  \ref{fig5}  refer to the first, third and sixth nodes of the chain, respectively.

\subsection{Amplifying on a frequency comb.}

We shall here demonstrate that the amplification can take place on a frequency comb. We shall in particular amplify a set of frequencies 
$\omega_k=\omega_0- k \Delta \omega$ with $k=0,1,2,..$; $\Delta \omega$ is positive and represents the relative distance between two consecutive frequency peaks. Reasoning as in the preceding section, we want to assign the volume of the second node so as to meet the condition $\frac{\omega_1}{\gamma_2} -\omega_0 \equiv - \Delta \omega$ which translates into:

\begin{equation}
 \label{V2}
V_2=\hat{V} \frac{1}{\omega_0/\Delta \omega -1} 
\end{equation}

where $\hat{V}=\frac{\omega_1}{\Delta \omega}V_1$.  Based on the same reasoning, we get for the other nodes the following recursive relation:

\begin{equation}
 \label{Vi}
V_i= \frac{V_{i-1}}{1- \frac{V_{i-1}}{\hat{V}}} 
\end{equation}

\begin{figure}
 \centering
   {\includegraphics[width=7.5cm]{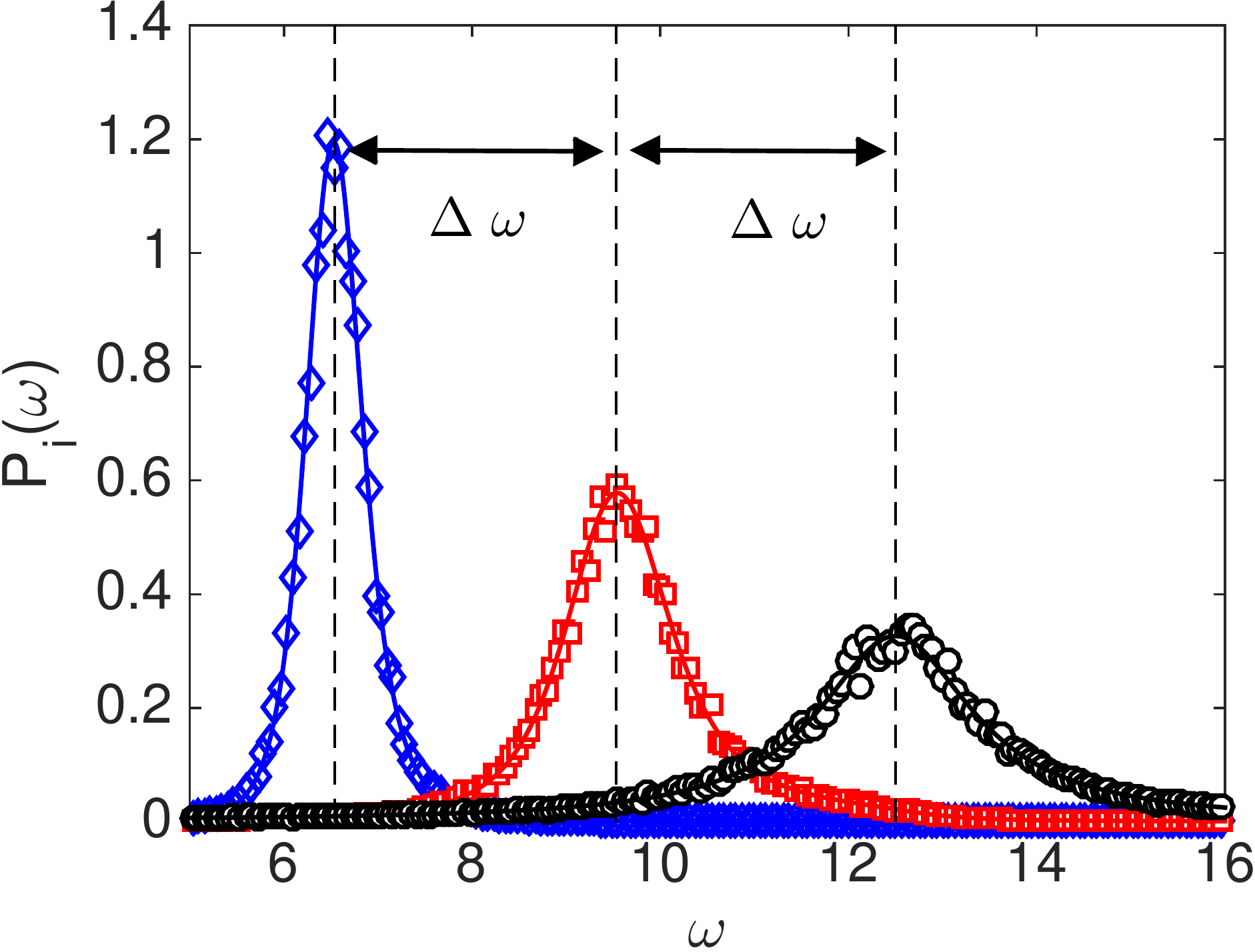}}
   \caption{Amplifying a frequency comb. Here, $\omega_k=\omega_0- k \Delta \omega$ with $k=0,1,2,..$. The positive quantity $\Delta \omega$ denotes the separation between two consecutive frequencies. The size of the volumes of the nodes are set as prescribed by equations (\ref{V2}) and (\ref{Vi}). The power spectra of fluctuations on different nodes (see text) are displayed. Symbols refer to direct simulations and the solid lines to the theory prediction. }
   \label{fig5}
  \end{figure}

As discussed in the preceding section, one can keep the volumes unchanged over a few consecutive nodes, before modulating their size as prescribed by formulae (\ref{V2}) and (\ref{Vi}), so to enhance the amplification power of the device. In Figure  \ref{fig5} we created a chain that implements the sequence of volumes $(V_1, V_2, V_2, V_3, V_3, V_3)$. The power spectra displayed in Figure  \ref{fig5}  refer to the first, third and sixth nodes, respectively.

\subsection{A consistent thermodynamic interpretation.}

Given the  probability density $P({\boldsymbol v},\tau)$ that obeys to the Fokker-Planck equation (\ref{FP}) we define the entropy
$S(\tau)$ as
\[
S(\tau)=-\int P({\boldsymbol v},\tau)\ln P({\boldsymbol v},\tau) d{\boldsymbol v} 
\] 
By deriving with respect to time $\tau$ the previous equation, one gets:
\[
\frac{dS}{d\tau}=-\int \frac{\partial P}{\partial\tau}\big(\ln P+1\big) d{\boldsymbol v}=\int\sum_i \frac{\partial I_i}{\partial v_i} \big(\ln P+1\big) d{\boldsymbol v}
\]
and, integrating by parts:
\[
\frac{dS}{d\tau}=-\sum_i \int I_i \frac{\partial}{\partial v_i}\ln P d{\boldsymbol v}
\]
By making use of the definition of the current ${\boldsymbol I}$, see main body of the paper, we write: 
\[
\frac{\partial}{\partial v_i}\ln P=\frac{2}{B_{ii}}A_i-\frac{2}{B_{ii}}\frac{I_i}{P}
\]
and finally: 

\[
\frac{dS}{d\tau}=\Pi_S-\Phi_S
\]

where

\begin{eqnarray}
{\Pi_S}&=&\sum_i \frac{2}{B_{ii}}\int \frac{I_i^2({\boldsymbol v},\tau)}{P({\boldsymbol v},\tau)} \\
{\Phi_S}&=&\sum_i \frac{2}{B_{ii}}\int A_i({\boldsymbol v})I_i({\boldsymbol v},\tau) d{\boldsymbol v} \nonumber
\end{eqnarray}

$\Pi_S$ is positive definite and can be interpreted as the production rate of entropy due to the non-conservative forces $A_i$. 
$\Phi_S$ can take in principle any sign. When $\Phi_S>0$, the entropy flows from the system to the environment. At equilibrium $I_i=0$, which implies $\Pi_S=\Phi_S=0$. A non trivial stationary solution exists which corresponds to setting $\Pi_S=\Phi_S \ne 0$. This is equivalent to imposing  $\sum \frac{\partial}{\partial v_i} I_i=0$, the condition of Fokker-Planck stationarity. The solution of the Fokker-Planck equation is hence interpreted as a dynamical balance between two opposing entropy fluxes. To quantify the entropy production $\Pi_S$, we can therefore estimate the antagonist contribution $\Phi_S$.

By making use of the definition of the current,  an performing an integration by parts, one gets:
\begin{equation}
\begin{split}
\Phi_S &=\sum_i\frac{2}{B_{ii}}\int A_i I_i d\boldsymbol v=\sum_i\frac{2}{B_{ii}}\int \big( A_i^2 P-\frac{B_{ii}}{2}A_i\frac{\partial}{\partial v_i}P)=\\
&= \sum_i\frac{2}{B_{ii}}\int \big( A_i^2 P+\frac{B_{ii}}{2}P\frac{\partial}{\partial v_i}A_i)=\\ 
&= \sum_i \big( \frac{2}{B_{ii}} <A_i^2 >+<\frac{\partial}{\partial v_i}A_i>\big)
\end{split}
\label{PhiNonLin}
\end{equation}

The above formula con be employed to determine the (non linear) entropy production rate $\Pi_S$ ($=\Phi_S$), displayed by the system in  stationary conditions. To gain analytical insight we can proceed with a direct estimate of $\Pi_S$ (and hence $\Phi_S$) that builds on the linear noise approximation. In this case we can write:

\[
\Phi_S= \sum_i \big( \frac{2}{{\mathcal B}_{ii}} <f_i^2 >+<\frac{\partial}{\partial \zeta_i}f_i>\big)
\]
where the non conservative force is now $f_i=(J\zeta)_i$.  
Recalling that:
\[
\sum_i <\frac{\partial}{\partial \zeta_i}f_i>=\sum_i <\sum_j J_{ij} \frac{\partial \zeta_j}{\partial \zeta_i}>=\sum_i <J_{ii}>=Tr(J)
\]
we can write: 
\[
\Phi_S=\sum_{i,j,k} \frac{2}{{\mathcal B}_{ii}} J_{ij}J_{ik}<\zeta_j\zeta_k>+Tr(J)
\]
Define then the correlation matrix $C_{ij}=<\zeta_i\zeta_j>$ and write
\begin{equation}
\begin{split}
\label{Phi}
\Phi_S &=2\sum_{i,j,k} \frac{1}{{\mathcal B}_{ii}} J_{ij}J_{ik}C_{jk}+Tr(J)=\\
&=2\sum_{i}\frac{1}{{\mathcal B}_{ii}}(JCJ^t)_{ii}+Tr(J)
\end{split}
\end{equation}

\begin{figure}
 \centering
   {\includegraphics[width=7.5cm]{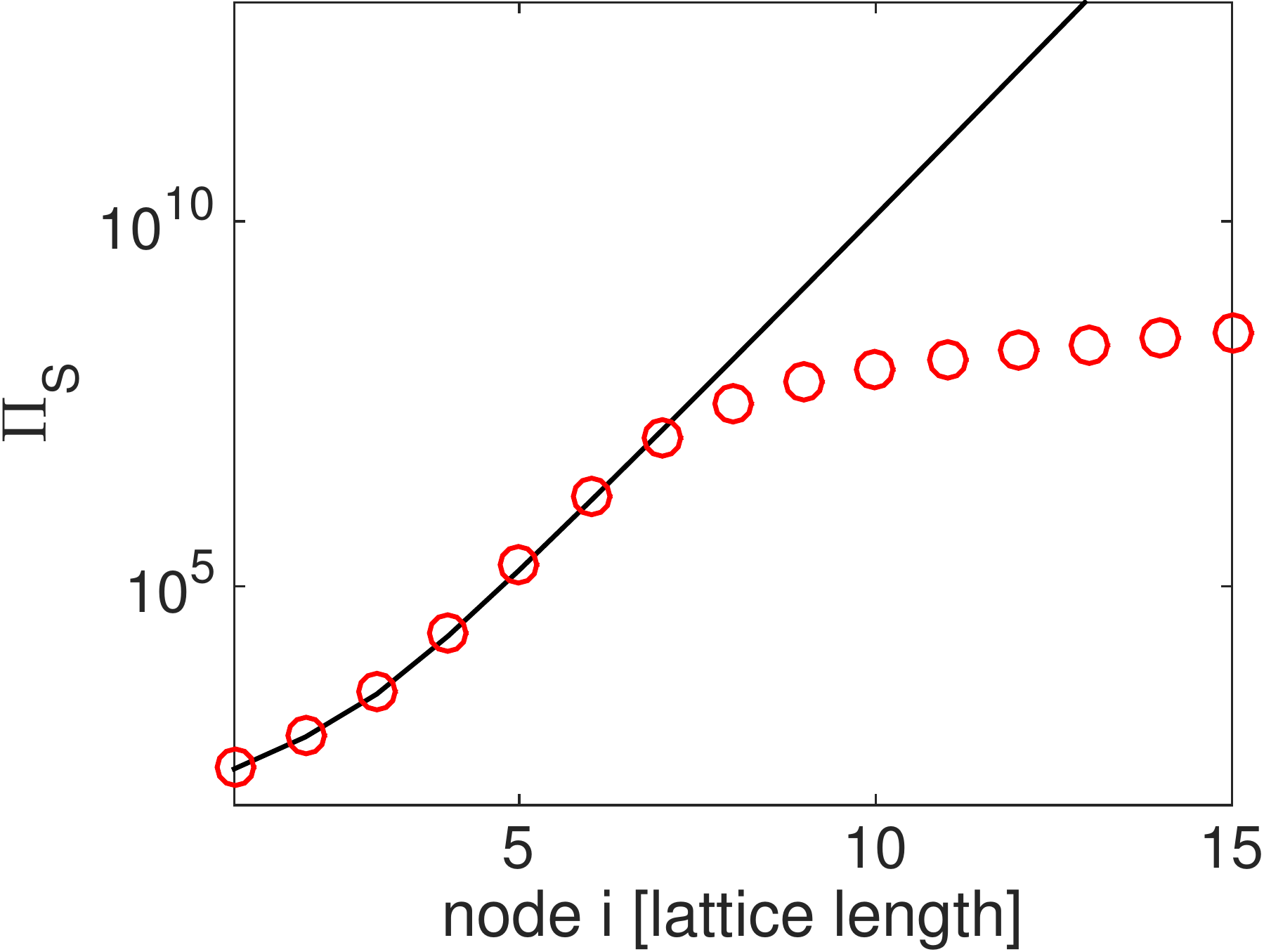}}
   \caption{ $\Pi_S$ is plotted (solid line) versus the lattice node. The solid line refers to the analytical estimate based on  linear noise approximation, see equation (\ref{Phi})). Symbols refer instead to the numerical estimate based on the fully non linear relation (\ref{PhiNonLin}). 
}
   \label{fig6}
  \end{figure}

In Figure \ref{fig6} the entropy production rate $\Pi_S$ ($=\Phi_S$, as given by formula (\ref{Phi})) is plotted (solid line) versus the lattice node, an indirect measure of the lattice length. As expected, $\Pi_S$ grows exponentially. Symbols refer instead to the a direct numerical characterization of $\Pi_S$, based on relation (\ref{PhiNonLin}). Non linear effects induce a cross-over towards a non exponential growth for the measured entropy production rate.

A cross-over towards a non exponential regime is eventually observed when non linearities become prominent, in complete agreement with the insight gained under a purely dynamical angle.

\subsection{On the validity of the Kramers-Moyal approximation: a numerical test.}

\begin{figure}
 \centering
   {\includegraphics[width=7.5cm]{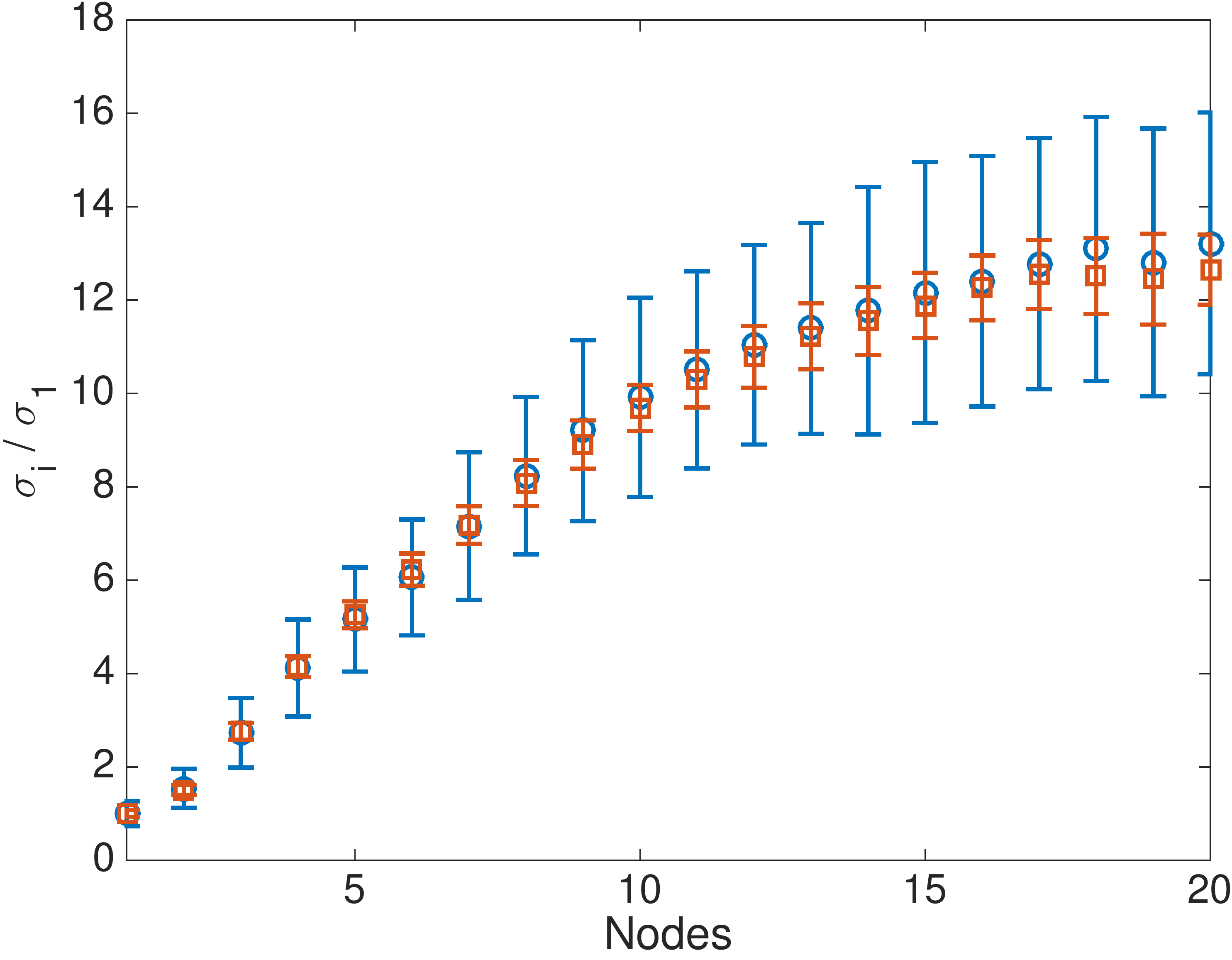}} 
   \caption{ $\sigma_i/\sigma_1$ vs. a progressive index that identifies the nodes location. Circles refer to Gillespie based simulations, squares to a direct integration of the non linear Langevin equations [Eqs. (1) in the main paper]. The error bars are obtained by averaging over different realizations of the stochastic dynamics. Here, $D=10$, $r=50$, $V = 20000$.}
   \label{fig7}
  \end{figure}

We here aim at testing the adequacy of the non linear Langevin equations, subject to multiplicative noise, as derived within the Kramers-Moyal picture. To this end we numerically evaluate  $\sigma_i/\sigma_1$, as defined in the main body of the paper, by using (i)
the Gillespie algorithm (which returns an exact description of the underlying master equation) and (ii) the non Linear equations (see main paper). The analysis is carried out for a sufficiently small volume amount, so that Gillespie based simulations are relatively unexpensive.  The comparison as drawn in Figure  \ref{fig7} certifies the accuracy of the Langevin representation of the dynamics.

\end{document}